# A Pathway to Efficient Simulations of Charge Density Waves in Transition Metal Dichalcogenides: A Case Study for TiSe$_2$

Li Yin[1,*], Hong Tang[1], Tom Berlijn[2], and Adrienn Ruzsinszky[1]

[1]*Department of Physics and Engineering Physics, Tulane University, New Orleans, Louisiana 70118, USA*

[2]*Center for Nanophase Materials Sciences, Oak Ridge National Laboratory, Oak Ridge, Tennessee 37831, USA*

---

[*]Author to whom all correspondence should be addressed.

E-mail: lyin2@tulane.edu

# ABSTRACT

Charge density waves (CDWs) in transition metal dichalcogenides are the subject of growing scientific interest due to their rich interplay with exotic phases of matter and their potential technological applications. Here, using density functional theory with advanced meta-generalized gradient approximations (meta-GGAs) and linear response time-dependent density functional theory (TDDFT) with state-of-the-art exchange-correlation kernels, we investigate the electronic, vibrational, and optical properties in $1T$-TiSe$_2$ with and without CDW. In both bulk and monolayer TiSe$_2$, the electronic bands and phonon dispersions in either normal (semi-metallic) or CDW (semiconducting) phase are described well via meta-GGAs, which separate the valence and conduction bands just as HSE06 does but with significantly more computational feasibility. Instead of the underestimated gap with standard exchange-correlation approximations and the overestimated gap with screened hybrid functional HSE06, the band gap of the monolayer TiSe$_2$ CDW phase calculated by the meta-GGA MVS (151 meV) is consistent with the angle-resolved photoemission spectroscopy (ARPES) gap of 153 meV measured at 10 K. In addition, the gap of bulk TiSe$_2$ CDW phase reaches 67 meV within the TASK approximation, close to the ARPES gap of 82 meV. Regarding excitations of many-body nature, for bulk TiSe$_2$ in normal and CDW phases, the experimentally observed humps of electron energy loss spectroscopy and plasmon peak are successfully reproduced in TDDFT, without an obvious kernel dependence. To unleash the full scientific and technological potential of CDWs in transition metal dichalcogenides, the chemical doping, heterostructure engineering, and pump-probe techniques are needed. Our study opens the door to simulating these complexities in CDW compounds from first principles by revealing meta-GGAs as an accurate low-cost alternative to HSE06.

# INTRODUCTION

Over the last two decades, transition metal dichalcogenides (TMDCs) have been the central focus in basic scientific studies and explorations of technological applications in the fields of electronics, magnetism, and optics[1–3]. Particularly, TMDCs are among the most representative materials testbeds for studying the interplay of long-range charge order, magnetic order, and superconductivity[4–8]. In the quasi two-dimensional $2H$-NbSe$_2$[9–11], multiband superconductivity and charge density wave (CDW) coexist. Bulk $1T$-TiSe$_2$ can also be superconducting with the CDW suppressed by either Cu intercalation[12] or pressure[4,13]. Although the intrinsic ferromagnetism is still under debate, the interplay between CDW and ferromagnetic order in $1T$-VSe$_2$ provides a fertile ground for achieving a Curie temperature higher than 330 K[3,14,15], which is prominently high among ferromagnetic transition temperatures in two-dimensional magnets. Given that CDWs often play a critical role in the formation of magnetic order or superconductivity, exploring the related ground and excited states becomes crucial.

The low-temperature properties of $1T$-TiSe$_2$ (a non-magnetic material) are a challenge to explanations. $1T$-TiSe$_2$ undergoes a CDW transition at the critical temperature of 202 K[13,16]. So far, both the origin of the CDW and the metallic/semi-metallic nature of $1T$-TiSe$_2$ are highly disputed. The origin of the CDW in $1T$-TiSe$_2$ has been explained mainly by two mechanisms. One is the excitonic effect due to electron-hole interaction[13,17], which is attributed to the proximity of valence and conduction bands and a competing Coulomb attraction of the electron and hole. Another possible mechanism is the electron-phonon coupling with a band-type Jahn-Teller instability[17,18]. The first mechanism can explain the semi-metal feature of $1T$-TiSe$_2$ reflected by transport data[1], although this mechanism can in other materials produce an excitonic insulator. However, according to the angle-resolved photoemission spectroscopy (ARPES) measurement, $1T$-TiSe$_2$ shows an indirect gap[19–21], which is as small as ~40 meV in the normal phase and ~80 meV in the CDW phase. Although such a gap feature matches the electron-phonon coupling mechanism, it still leaves the ambiguity of $1T$-TiSe$_2$ being semiconducting given its semi-metallic transport feature. Besides, both the Lindhard susceptibility and nesting function in $1T$-TiSe$_2$ don't change as the CDW is suppressed by pressure[4]. So, the Fermi surface nesting mechanism is out of the picture in TiSe$_2$. Irrespective of the driving mechanism, both electron-hole interaction and band-type Jahn-Teller transitions demonstrate the highly sensitive electronic properties in TiSe$_2$. Furthermore, previous hybrid functional calculations -including an adjusted Heyd-Scuseria-Ernzerhof (HSE06)[22,23] for TiSe$_2$- show that the screened exchange interaction can capture the electron-electron interaction to separate the *d* and *p* states and stabilize the CDW phase[24]. Therefore, the CDW in TiSe$_2$ is assumed to be purely or strongly driven by electronic instability, and TiSe$_2$ becomes a perfect candidate for digging into the role of exchange and correlation interactions in CDW. However, the hybrid functional computations are expensive, which hamper the theoretical computations of complexities induced by chemical doping, heterostructure engineering and external perturbations on ultrafast timescales of CDWs in TMDCs[25–27]. To overcome these challenges, different functionals are needed that are computationally less expensive than the hybrid functional yet retain their accuracy in terms of describing CDWs in TMDCs. We intend to employ the low-cost but advanced meta-generalized gradient (meta-GGA) density functional approximations -the best possible way at the semilocal level- to approach the exchange-correlation energy in TiSe$_2$[28,29]. The extra kinetic energy density ingredient or Laplacian dependence of the local electron density makes meta-GGAs flexibly satisfy different exact constraints[30,31].

Accompanying the unusual trend in the change of resistivity with temperature[1,32], another interesting behavior of 1T TiSe$_2$ is the dramatic plasmon response with temperature, as shown from optical reflectivity experiments[1] and the high resolution momentum resolved electron energy loss spectroscopy (HREELS) measured at the zero momentum transfer[33,34], where at low temperatures of ~10-35 K, a very defined sharp plasmon peak resides at about 50 meV, while this plasmon peak becomes weaker and broader with the increase in temperature and is completely undiscernible above the CDW transition temperature of 200 K. In regard of the origin of the CDW transition in 1T TiSe$_2$, there are still debates over the plasmon softening[34] and the interband transition induced Landau damping to plasmons with finite momenta[33]. Either way, the strong tunability of the plasmon in the semi-metal 1T TiSe$_2$ or small-gap semiconductor can play an instrumental role in plasmonic applications.

In this work, employing the density functional theory (DFT) with low-cost advanced meta-GGAs and linear response time dependent density functional theory (lr-TDDFT), we investigate the electronic, vibrational, and optical properties for bulk and monolayer 1*T*-TiSe$_2$. We find that the electronic bands and phonon dispersions in bulk and monolayer TiSe$_2$ are accurately described by meta-GGAs, in either normal or CDW phase. Different from the underestimated gap by standard exchange-correlation approximations or the overestimated band gap by hybrid functional

HSE06, the MVS-based band gap for monolayer TiSe$_2$ and the TASK-based gap for bulk TiSe$_2$ in the CDW phase match the experimental gaps. Moreover, the humps of EELS and the plasmon peak for bulk TiSe$_2$ in normal and CDW phases are successfully reproduced by TDDFT without obvious kernel dependence. Our research paves the way for unlocking the full scientific and technological potential of CDWs in TMDCs.

## METHODS

Ab initio calculations were performed with the Perdew-Burke-Ernzerhof GGA[35] and advanced meta-GGAs[31,36,37], implemented in the Vienna Ab-initio Simulation Package (VASP)[38,39]. The projector augmented wave pseudo-potentials[40,41] and energy cutoff of 320 eV are utilized. The meta-GGA made very simple (MVS)[36], the strongly constrained and appropriately normed (SCAN)[31], the regularized-restored SCAN (r$^2$SCAN)[42,43], TASK, and modified TASK (mTASK)[44,45] are utilized in this work, where the first four approximations are constructed for excellent accuracy for ground state properties, and the last two approximations are built upon an exchange enhancement factor with enhanced spatial nonlocality. The valence electron configurations of $3d^34s^1$ and $5s^25p^4$ are considered in Ti and Se atoms, respectively.

The Brillouin zone is sampled with a Γ-centered 18×18×11 $k$ point mesh in bulk TiSe$_2$ and a 18×18×1 $k$ point mesh in monolayer TiSe$_2$. The vacuum region of 23 Å is employed in the $c$ direction for simulating the two-dimensional characteristics of monolayer TiSe$_2$. The convergence criteria are 10$^{-6}$ eV for the energy and 0.001 eV/Å for the atomic forces, respectively. The lattice constants of bulk TiSe$_2$ are fixed at the experimental values of $a$=3.540 Å and $c$= 6.007 Å[46,47]. The lattice constants of monolayer TiSe$_2$ are fixed at the experimental values of $a$=3.538 Å[48]. The atomic coordinates in bulk and monolayer TiSe$_2$ are fully relaxed within the PBE-GGA approximation. The relaxed z-directional distance between Ti and Se layers in bulk TiSe$_2$ is 1.535 Å, which is close to the experimental value of 1.532 Å[47]. This z-directional distance between Ti and Se layers is relaxed to 1.546 Å in monolayer TiSe$_2$. Interatomic force constants are calculated by the supercell approach implemented in Phonopy[49,50], with the energy cutoff of 550 eV and energy convergence criteria of 10$^{-6}$ eV. For bulk (monolayer) TiSe$_2$, the supercell 4×4×4 (8×8×1) and Γ-centered $k$-mesh 4×4×2 (5×5×1) are used for phonon dispersions.

The 2×2×2 CDW structure in bulk TiSe$_2$ is relaxed with the initial displacement of **d**$_{3L}$ pattern and the experimental ratio of δTi/δSe=3[24,51]. On the basis of the DFT calculations in the 2×2×2

CDW structure, the projected Wannier function method is further used to get the tight-binding Hamiltonian[52]. Then, using the eigenvalues and eigenvectors of the output Hamiltonian, we unfold the bands and Fermi surface of the 2×2×2 CDW supercell into the Brillouin zone of the primitive cell with a proper spectral weight[53]. The Wannier-function based Hamiltonian enables the unfolding method to be performed on a dense k grid at low computation cost. The Ti $d$ and Se $p$ orbitals are projected, without the maximal localization in Wannier90[54]. The crystal structure is visualized using the VESTA 3.4.7 code[55]. The electronic structures are plotted using the Gnuplot 4.6 code[56].

The electron energy loss spectroscopy (EELS) calculations for both normal and CDW phases of 1T TiSe$_2$ are performed with the lr-TDDFT method. The flavor of the TDDFT exchange-correlation kernels used in the calculations include the recently constructed frequency-dependent modified Constantin-Pitarke 2007 (MCP07)[57], the fully adiabatic local density approximation (ALDA)[58], and the long-range-corrected (LRC) designed for the optical absorption spectra of small and medium-gap semiconductors[59]. The random phase approximation (RPA)[60] and independent particle (IP) methods[61,62] are also used for comparison. All the calculations are for zero momentum transfers (i. e. q=0), and the intraband transitions for the metallic phase are included through the Drude term[63], which is expressed as $\frac{\omega_{pl}^2}{\omega(\omega+i\gamma)}$, where the real part of the plasmon frequency $\omega_{pl}$ is the unscreened plasmon frequency $\omega_{pl} = \sqrt{4\pi n/m^*}$ in atomic Hartree unit, and $n$ and $m^*$ are the density and the effective mass of the charge carrier, respectively. $\gamma$ is the imaginary part of the plasmon frequency and represents the scattering rate or relaxation time of the charge carrier. The experimental results show a large reduction both in the charge carrier density and the effective mass from room temperature to low temperature[64]. We estimate that the unscreened plasmon frequencies for the normal and the low temperature (CDW) phase are similar in values and set $\omega_{pl}$ to 0.5 eV, and this leads to the calculated positions of the plasmon peak for both phases consistent with experimental results[1,33]. The imaginary part $\gamma$ is set to the typical value of 0.1 eV for the low temperature phase[33], while it is set to 11 eV for the normal phase to be consistent with the experimentally observed large reduction in the charge carrier mobility from low temperatures to high temperatures[64,65].

## RESULTS AND DISCUSSION

**Ground state electronic and vibrational properties in TiSe$_2$.** In bulk and monolayer TiSe$_2$, the 2×2×2 and 2×2×1 supercells are distorted in the **d**$_{3L}$ pattern, then form the CDW phases, as displayed in Figs. 1a and b. We calculated the energy difference as a function of atomic displacement with PBE and different meta-GGA approximations. As shown in Figs. 1c and d, except for the mTASK in monolayer, almost all the meta-GGA approximations provide a stronger negative curvature and corresponding energy gain than that with PBE, in either the bulk or monolayer TiSe$_2$. Especially, the MVS approximation produces the strongest energy gain. We further calculate the band structures of the undistorted bulk and monolayer TiSe$_2$, which are shown in Figs. 1e-h, Figs. S1 and S2 of the supplement. The MVS energy gain and structural distortion are close to our reference values[24]. The band structures of both bulk and monolayer TiSe$_2$ are described well within various meta-GGAs. The states of the valence band maximum (VBM) and conduction band minimum (CBM) around the Fermi level primarily consist of Se 4$p$ orbital and Ti 3$d$ orbital, respectively. The band shapes that are electron-like for the conduction band at the M/L points and hole-like for the valence band at the Γ/A points are reproduced. Moreover, as shown in Figs. 1g, h, S1, and S2, for bulk or monolayer TiSe$_2$, meta-GGA approximations separate the valence and conduction bands around the Γ/A points, decreasing the metallicity of TiSe$_2$, just as the hybrid functional HSE06 does[24]. Also, we note that, along with the separated valence and conduction bands, the $d$-$p$ orbital hybridization from Γ to A is weakened, as compared with PBE.

Among the considered meta-GGA approximations, the separation between valence and conduction bands in TASK or mTASK approximation is stronger than that in SCAN, r$^2$SCAN, or MVS approximation as shown in Figs. 1g, h, S1, and S2, which applies for both bulk and monolayer TiSe$_2$. This phenomenon is rooted in the different construction strategies of meta-GGA approximations. Generally, the exchange-correlation energy in meta-GGAs is defined by the electron density, local gradient of the densities, and the kinetic energy density constructed from the occupied Kohn-Sham orbitals[31,36,37]. As compared with GGAs, the kinetic energy density is added in meta-GGA to distinguish the iso-orbital and nearly-uniform electron gas regions. Then, an ingredient $\alpha = \frac{\tau - \tau^W}{\tau^{unif}}$ is introduced to characterize the dimensionless derivation from a single orbital shape, where $\tau^W = \frac{|\nabla n|^2}{8n}$ is the von Weizsäcker kinetic energy density and $\tau^{unif} = \frac{3(3\pi^2)^{2/3} n^{5/3}}{10}$ is the kinetic energy density limit of uniform electron gas. Then, the exchange enhancement factor $F_x$ is further defined by the ingredient $\alpha$ and the reduced density gradient $s =$ $\frac{|\nabla n|}{2(3\pi^2)^{1/3} n^{4/3}}$ [36]. Meta-GGA approximations show a dichotomy with respect to $F_x$. One construction strategy of $F_x$ can lead to meta-GGA functionals with excellent accuracy for ground state properties[66]. The other construction strategy emphasizes nonlocality in meta-GGAs altering the slope of the exchange energy density with respect to the kinetic energy density ingredient. The former strategy is the basis of the construction scheme of MVS, SCAN, and r$^2$SCAN approximations. TASK and mTASK are built upon an exchange enhancement factor with enhanced nonlocality. Unlike directly-modelled meta-GGA potentials[67], the effective potentials for our meta-GGAs are properly functional derivatives. The reduced screening in two-dimensional materials raises the bar for more nonlocal meta-GGA approximations when computing fundamental band gaps and band structures. The modified TASK (mTASK) was designed to fill this gap by creating more nonlocality in the exchange component. With a targeted alteration of the original TASK functional the slope $\frac{\partial F}{\partial \alpha}$ can be made more negative so that mTASK matches the screening in low-dimensional materials.

In Figs. 1c and d, the MVS approximation produces the strongest energy gain, and the one closest to that of the HSE(17,0) long-range hybrid with 17% of exact exchange at all ranges[24], which can also be regarded as a PBE global hybrid. Considering the electronic structure, both TASK and mTASK approximations can strongly separate the valence and conduction band. However, the TASK approximation exhibits better energy gain for CDW instability than the mTASK, demonstrating that TASK is a good choice for balancing the CDW instability and electronic structure in the normal phase. In addition, we calculated the phonon dispersion of bulk and monolayer TiSe$_2$ in the normal phase. As displayed in Fig. 2, bulk and monolayer TiSe$_2$ is dynamically unstable at the L and M points in PBE, MVS, or TASK approximations, which correspond to the 2×2×2 and 2×2×1 supercells in CDW phase. Particularly, in the MVS-based monolayer TiSe$_2$ displayed in Fig. 2d, although the first phonon branch with imaginary frequency is found near the Γ point, the strongest imaginary frequency valley is still located at the M point, corresponding to 2×2×1 supercell.

We stress that within this work, the phonon dispersions of the relaxed structures are based on the experimental lattice constants. We have also fully relaxed the bulk and monolayer TiSe$_2$ with the Grime D2 van der Waals correction[68] and different approximations (please see the Table S1 of the supplement), where the SCAN- and r$^2$SCAN-relaxed lattice constants are close to the experimental values[46–48] and comparable to HSE06 and HSE(17,0) results[24].

Overall, given the energy difference curves, band structures, and phonon dispersions, meta-GGAs, especially MVS and TASK, are proper exchange-correlation approximations to describe the bulk and monolayer TiSe$_2$, which largely supersede the accuracy of PBE and competes in accuracy with HSE06 with less computational resources.

**Electronic structures in TiSe$_2$ with CDW.** Based on the well-described TiSe$_2$ in the undistorted normal phase, we further investigate the electronic structure of the distorted CDW phase. For comparison, the band structure and energy surface are calculated within PBE, MVS, and TASK approximations. First, the bands calculated by Wannier functions and DFT are consistent, as illustrated in Fig. S3 of the supplement. The corresponding densities of states in bulk and monolayer TiSe$_2$ are also provided in Fig. S3. Then, we employ the Wannier-function-analyzed Hamiltonian to unfold the electronic structure of distorted CDW supercell, as shown in Fig. 3. For bulk TiSe$_2$, the valence bands around the Γ point move downward slightly as the phase changes from normal to CDW, which applies to either the PBE, MVS, or TASK approximation. Then, just like the normal case, the MVS and TASK approximations separate the valence and conduction bands, where the Γ-focused VBM and M-focused CBM are not overlapping anymore, as shown in Figs. 3c and e. A small band gap of 67 meV appears in bulk TiSe$_2$ within the TASK approximation, reasonably close to the ARPES-measured gap of 82 meV at 10 K[19]. In monolayer TiSe$_2$, a band gap is opened in the CDW phase, as displayed in Figs. 3b, d and f. The band gap reaches 67 meV with PBE, 151 meV with MVS, and 297 meV with TASK, where the MVS gap is very close to the ARPES-measured gap of 153 meV at 10 K[19]. For reference, the HSE06-calculated gap in monolayer TiSe$_2$ with CDW is 330 meV[19]. Besides, the shapes of MVS- and TASK-calculated energy surfaces at $E_F$-1 eV for bulk TiSe$_2$ highly resemble the ARPES-measured constant-energy-contour maps. All of these features demonstrate that MVS and TASK approximations work well in both undistorted and distorted TiSe$_2$, either in its bulk or monolayer form. These results can be the basis of ab-initio inputs for some beyond DFT calculations or machine learning applications in future.

One may also note that the top valence band splitting around the Γ or M point is revealed by ARPES, while that disappears in the calculated band structures of bulk and monolayer TiSe$_2$. Such band splitting can be corrected by spin-orbit coupling (SOC), which is verified by the TASK+SOC example as supplemented in Fig. S4. At the same time, the TASK-based band gap of monolayer

TiSe$_2$ in the CDW phase is decreased by 25 meV, as shown in Figs. 3f and S4b. We emphasize that correcting this band splitting is necessary for analyzing the optical properties of TiSe$_2$, such as the EELS or plasmon dispersion, and revealing if the excitonic effect is the origin of CDW transition in TiSe$_2$.

**Optical properties of bulk TiSe$_2$ with and without CDW.** In the above sections, we studied the electronic and vibrational properties of TiSe$_2$, and opened the CDW gap successfully with meta-GGA functionals. However, until now, the conflicted transport data and ARPES-measured gap leave some ambiguity about the semi-metal or semiconductor nature of TiSe$_2$. TiSe$_2$ is consistently reported as a specific material whose CDW phase exhibits a band gap opening and metallic conduction simultaneously. Metallic phases can be characterized by sharp plasmon peaks as explored in this manuscript.

Still, the experiments show mixed results on the semi-metal or semiconductor nature of TiSe$_2$. Especially, the normal state of TiSe$_2$ can be semi-metal or small gap semiconductor, depending on the preparation conditions[16,69–71]. However, all prepared samples show a CDW transition below a certain CDW transition temperature and the CDW gap can be detected by ARPES measurements. The magnetoresistance and quantum oscillation measurements show both electron and hole pockets in the normal state, but only an electron pocket in the low temperature CDW state, with most of the charge carriers being gapped out[64]. Interestingly, the process is accompanied by a dramatic plasmon response, which shows a prominent sharp peak in the low-temperature state that is completely suppressed in the high temperature normal state[1,33,34]. Here we show the intraband transition nature of this plasmon peak through TDDFT calculations with a Drude term correction. Our simulations are based on the semi-metallic ground state for the normal phase, and, for the CDW phase, both a metallic ground state and an electron-doped gapped ground state are considered. Those considerations are approximately consistent with experimentally described pictures[64].

Inspired by the dramatic plasmon response experimentally measured in bulk TiSe$_2$, we theoretically simulate the EELS for bulk TiSe$_2$ with and without a band gap. We first check the effect of SOC on the calculated EELS of TiSe$_2$, where the Drude term is switched off[63]. For the normal phase, the calculated EELS with SOC shows humps around 1000-2000 cm$^{-1}$, which is approximately consistent with the experimental result, while the calculated EELS without SOC shows no humps around this energy range. For the CDW phase, however, both cases with and without SOC show humps around 4000 cm$^{-1}$, approximately consistent with the experiment. The comparison of EELS curves can be found in Fig. S5. Note that the above comparisons are based on semi-metallic or metallic ground states for both the normal and the CDW phases. For the gapped CDW phase, the SOC effect appreciably splits the top of the valence bands, and hence calculations with SOC are more realistic for optical calculations. In the following calculations, the SOC effect is included.

The calculated EELS of 1T TiSe$_2$ for both the normal and CDW phases are shown in Fig. 4. For the normal phase (at room temperature), the ground state is calculated with PBE+SOC and is semi-metallic. As shown in Fig. 4a, the hump around 1000 cm$^{-1}$ in EELS is reproduced by the MCP07 exchange-correlation kernel and IP calculations within lr-TDDFT[57,61,62], although the height of the hump is slightly overestimated by calculations. Fig. 4d shows the comparison of EELS results for the normal phase calculated with MCP07[57], RPA[60], LRC and ALDA kernels (with zero exchange-correlation kernel for RPA)[58,59]. The EELS curves calculated with MCP07 and RPA are very close to each other, and both are close to and resemble the independent particle (IP) curve in this energy range. The LRC (with the parameter $\alpha$=0.43 eV) and ALDA curves are slightly lower than those of MCP07 and RPA in the range of about 500-8000 cm$^{-1}$.

For the CDW phase, the EELS is calculated for both metallic and semiconducting ground states. The metallic ground state is calculated with PBE+SOC, and the related EELS results are shown in Figs. 4b and e. MCP07 and IP kernels give the plasmon peak at 400 cm$^{-1}$ as shown in Fig. 4b, which is close to the experimental value 347 cm$^{-1}$ at 10K as shown in Fig. 4c. Fig. 4e shows the comparison of the calculated EELS from MCP07, RPA, ALDA, and LRC for the CDW phase based on a metallic ground state. The four methods all show the sharp plasmon peak at 400 cm$^{-1}$. The heights of the peaks from MCP07, RPA and ALDA are almost the same and they are close to that of IP. The peak height of LRC at 400 cm$^{-1}$ is slightly lower than those of MCP07, RPA, ALDA and IP. A slightly different trend is seen in the range of 3000-10000 cm$^{-1}$, the range of higher energy interband transition, where the curves of MCP07 and RPA are close to each other and both are slightly higher than those of ALDA, LRC and IP.

The gapped CDW phase ground state is calculated with PBE+SOC+$U$ with $U$=3.5 eV, which gives a gap of 80.7 meV. The corresponding EELS are displayed in Fig. S6 of the supplement. Fig. S6a shows the results of EELS calculated from the gapped CDW ground state. Fig. S6b displays the same result with a smaller vertical scale. Since the transport experiment shows a net charge carrier density at low temperatures, we incorporate an electron doping level of $1.8 \times 10^{25}\ m^{-3}$, according to the experimental results[64]. The Drude term parameter $\omega_{pl}$ is set to 0.27 eV, which is close to the value evaluated from experimental measured data[64]. The calculated plasmon peak height (~0.5-0.6) is larger than the experimental one (~0.1). However, the peak position and the features around 3000-6000 cm$^{-1}$ are consistent with the experimental ones. The results from other TDDFT kernels are also similar and can be found in Figs. S6c and S6d. It has been found that the calculation of EELS based on the gapped CDW ground state also reproduces the experimental result.

Overall, for the normal phase and in the relatively low energy range of 0-1.3 eV, MCP07, RPA and IP show very similar EELS curves and more enhanced interband transition humps than ALDA and LRC do. For the CDW phase, all the methods considered here produce the sharp plasmon peak at 400 cm$^{-1}$, and MCP07 and RPA produce slightly enhanced humps in the slightly higher interband transition range of 3000-10000 cm$^{-1}$ than ALDA, LRC and IP do.

Approximately, MCP07, ALDA, and LRC all give EELS curves similar to the one from RPA, suggesting that the kernel effect is relatively weak for the optical calculation in the relatively low energy range of 0-1.3 eV in 1T TiSe$_2$. Both the intraband and interband transitions are important for the optical property features (EELS here). To further investigate the damping effect to the plasmon peak in the normal phase, we calculate the EELS with different $\gamma$ (the imaginary part of the plasmon frequency) in the Drude term, see Fig. S7. The results show that the plasmon peak can be effectively damped (or suppressed) only at large $\gamma$ values, indicating that the strong plasmon damping in the normal phase cannot be achieved by the interband damping channel only, and should be contributed by the intraband and/or the coupled intraband-interband damping effects.

The unscreened plasmon frequency estimated from the experimental measured parameters[64,72] is in the range of 0.18-0.27 eV, which is less than 0.5 eV, the value used in the calculations here. The difference may be attributed to the complexity in evaluations of the averaged total effective mass of the charge carriers involved in the plasmon mode, such as orientation and high anisotropy of electron pockets and heavy mass part of the bands around the Fermi level[72]. The calculated EELS of the CDW phase with $\omega_{pl}^{unscreened}$=0.27 and $\gamma$=0.1 eV based on the metallic ground state shows overall similar features, except for a redshift of the plasmon peak, as illustrated in Fig. S8 of the supplement.

The calculated results allow us to estimate the screening effects of the interband transitions to the plasmon mode in TiSe$_2$. The screening effect can be approximated with the high frequency dielectric constant $\varepsilon_\infty$ and $\omega_{pl}^{screened} = \omega_{pl}^{unscreened}/\sqrt{\varepsilon_\infty}$[33], where $\omega_{pl}^{unscreened}$ is the unscreened plasmon frequency (real part), which is set to 0.5 eV. The screened plasmon frequency (real part) $\omega_{pl}^{screened}$ of the normal phase is determined to be 809.3 cm$^{-1}$ from Fig. 4a, and the value for the CDW phase is 400.5 cm$^{-1}$ for Fig. 4b. This leads to the value of the relative dielectric constant $\varepsilon_\infty$ of 24.8 and 101.4 for the normal and CDW phases, respectively, indicating a stronger (or enhanced) interband transition screening effect to the plasmon mode in the CDW phase.

We have demonstrated that the ground state in TiSe$_2$ can be well described by constraint-based exchange-correlation functionals (meta-GGAs), and the excited states including humps of EELS and plasmon peaks can be reproduced by lr-TDDFT. However, even with these results, we still do not know if the excitonic effect is the origin of the CDW in TiSe$_2$. The small indirect gap and sharp plasmon peak corroborate an exciton insulator-driven mechanism but we stress that some physics of exciton insulators such as soft plasmons remains to be shown, possibly even by lr-TDDFT[34,57]. In the future, we will conduct more detailed simulations for excited properties, such as the exciton energy, momentum-resolved EELS, and plasmon dispersion, to clarify the excitonic effect in TiSe$_2$.

## CONCLUSION

In summary, using DFT with advanced meta-GGAs and lr-TDDFT with various exchange-correlation kernels, we closely approach the ground and excited states in 1*T*-TiSe$_2$ with/without CDW, including the electronic, vibrational, and optical properties. We establish the meta-GGA approximations (especially MVS and TASK) as computationally more appealing alternatives to HSE06 and related hybrids for the ground state of the CDW phase (using experimental lattice constants). In bulk and monolayer TiSe$_2$, the electronic bands and phonon dispersions in either normal or CDW phases are described well via meta-GGAs, which separate the valence and conduction bands just as the more expensive adjusted hybrid HSE06 does. Especially, the MVS-based band gap (151 meV) of monolayer TiSe$_2$ and TASK-based gap (67 meV) of bulk TiSe$_2$ in CDW phases are close to the ARPES values, which is better than the underestimated gap with PBE and overestimated gap with screened hybrid functional HSE06. Moreover, we calculated the EELS for bulk TiSe$_2$ in normal and CDW phases, where the humps of EELS and plasmon peak are successfully reproduced by TDDFT without obvious dependence on the exchange-correlation kernel, as one can expect in the plasmon-dominated wavevector regime[73]. These results provide a low-cost alternative to hybrid functionals for investigation of the sensitive properties of CDWs in TiSe$_2$ and other TMDCs, thereby stimulating a full exploration of their scientific and technological potential.

## ACKNOWLEDGEMENTS

This work was supported by the donors of ACS Petroleum Research Fund under New Directions Grant 65973-ND10. A.R. served as Principal Investigator on ACS PRF 65973-ND10 that provided support for H.T. L.Y. and A.R. acknowledge support from Tulane University's startup fund, which also supports L.Y. This research includes calculations carried out on HPC resources supported in part by the National Science Foundation through major research instrumentation grant number 1625061 and by the US Army Research Laboratory under contract number W911NF-16-2-0189. This research was supported in part by the high performance computing (HPC) resources and services provided by Information Technology at Tulane University, New Orleans, LA. Part of this research (T.B.) was conducted at the Center for Nanophase Materials Sciences, which is a DOE Office of Science User Facility.

# FIGURE CAPTIONS

**Fig. 1 Meta-GGA approximations on bulk and monolayer TiSe$_2$.** (**a**) The real and reciprocal lattice structures of TiSe$_2$ primitive cell (**a1**, **a2**, **b1**, and **b2**) and 2×2×2 supercell (**A1**, **A2**, **B1**, and **B2**). (**b**) The **d**$_{3L}$ pattern in distorted TiSe$_2$ CDW structure, displayed with δTi=0.3 Å. (**c**, **d**) The variation of energy difference with the displacements of Ti atoms in bulk and monolayer TiSe$_2$. The case of δTi=0 Å denotes the energy of undistorted normal phase calculated in a specific approximation. The atom-characterized band structures of undistorted (**e**, **f**) bulk and (**g**, **h**) monolayer TiSe$_2$ with PBE and MVS approximations, respectively. The red (blue) circles represent the Ti 3$d$ (Se 4$p$) character. The Fermi level is at 0 eV. The undistorted structures are semi-metallic. For the band structures of the distorted semi-conducting structures, see Fig. 3.

**Fig. 2 Vibration properties of TiSe$_2$ in normal phase.** The phonon dispersion of undistorted bulk and monolayer TiSe$_2$ with (**a**, **b**) PBE, (**c**, **d**) MVS, and (**e**, **f**) TASK approximations, respectively.

**Fig. 3 Electronic structures of TiSe$_2$ in CDW phase.** The unfolded band structures and energy surfaces for bulk and monolayer TiSe$_2$ in relaxed CDW phase with (**a**, **b**) PBE, (**c**, **d**) MVS and (**e**, **f**) TASK approximations, respectively. The red (blue) color denotes Ti 3$d$ (Se 4$p$) character. The displayed energy surface is located at $E_F$-1 eV. The Fermi level $E_F$ is at 0 eV.

**Fig. 4 Calculated EELS of bulk TiSe$_2$ solid in the normal and the CDW phase.** The EELS calculated with the MCP07 kernel and the independent particle (IP) approximation for (**a**) normal and (**b**) CDW phases. (**c**) The experimental EELS. Adapted with permission from Ref. 1. Copyright 2007, American Physical Society. (**d**) and (**e**) comparisons of the calculated EELS with different TDDFT kernels for the normal and CDW phases, respectively.

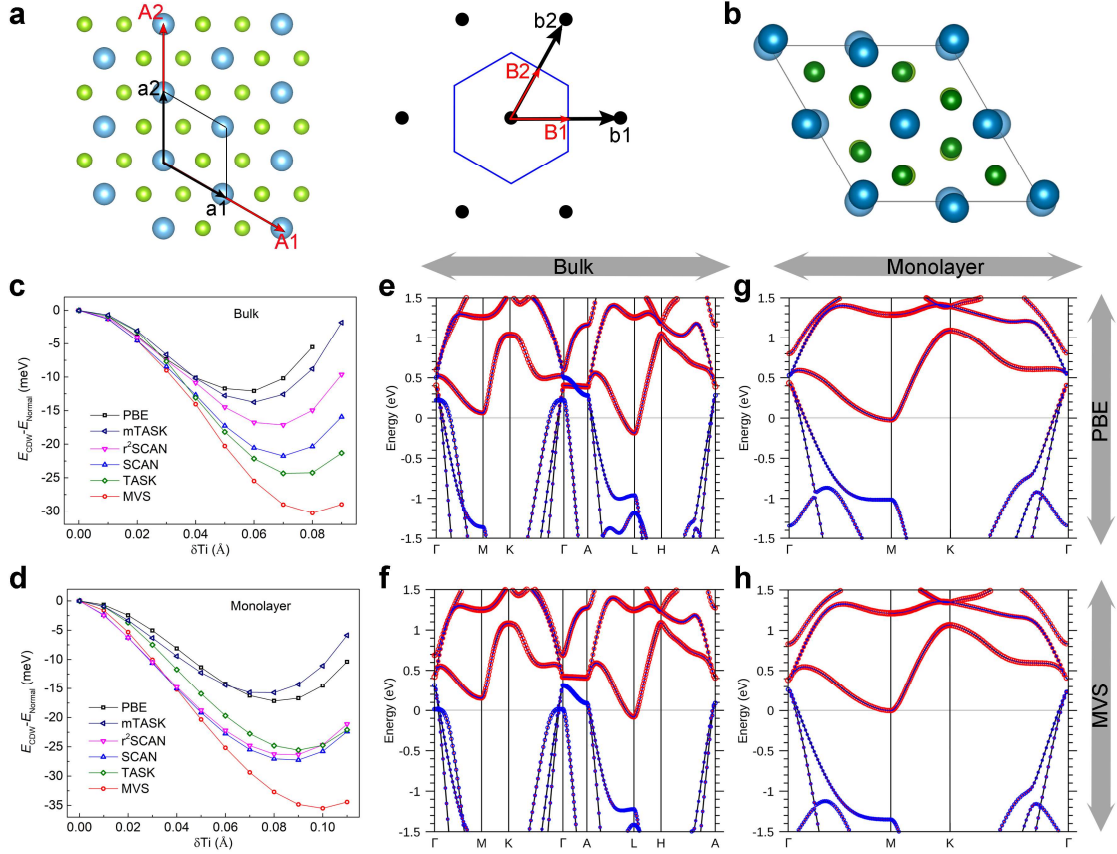

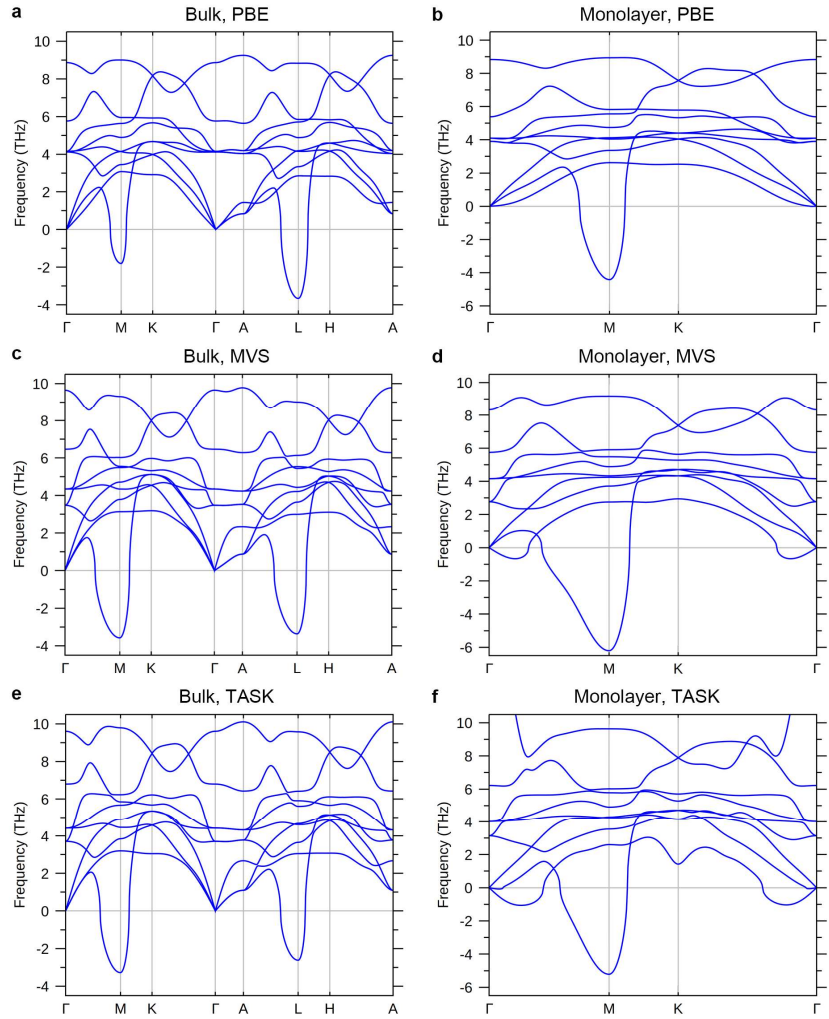

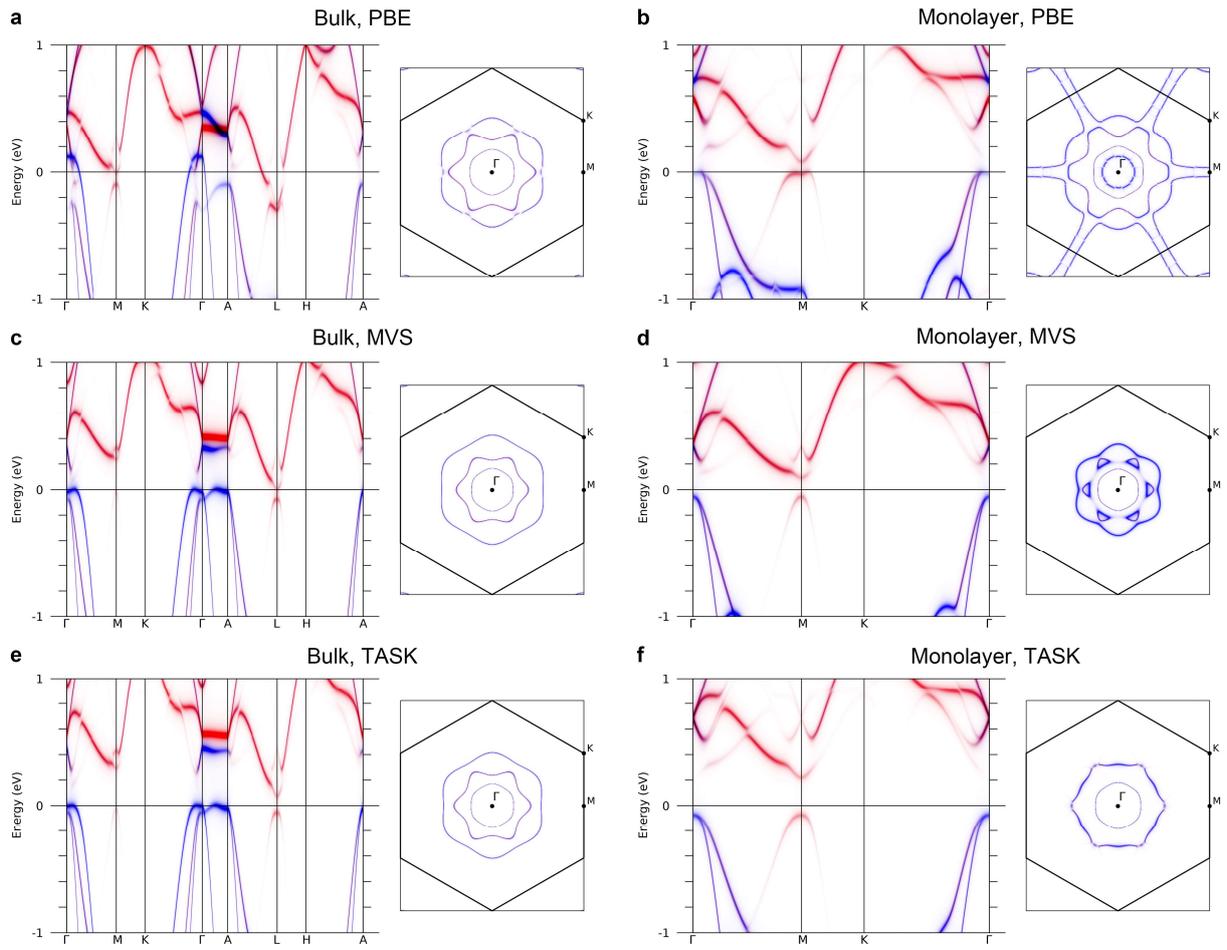

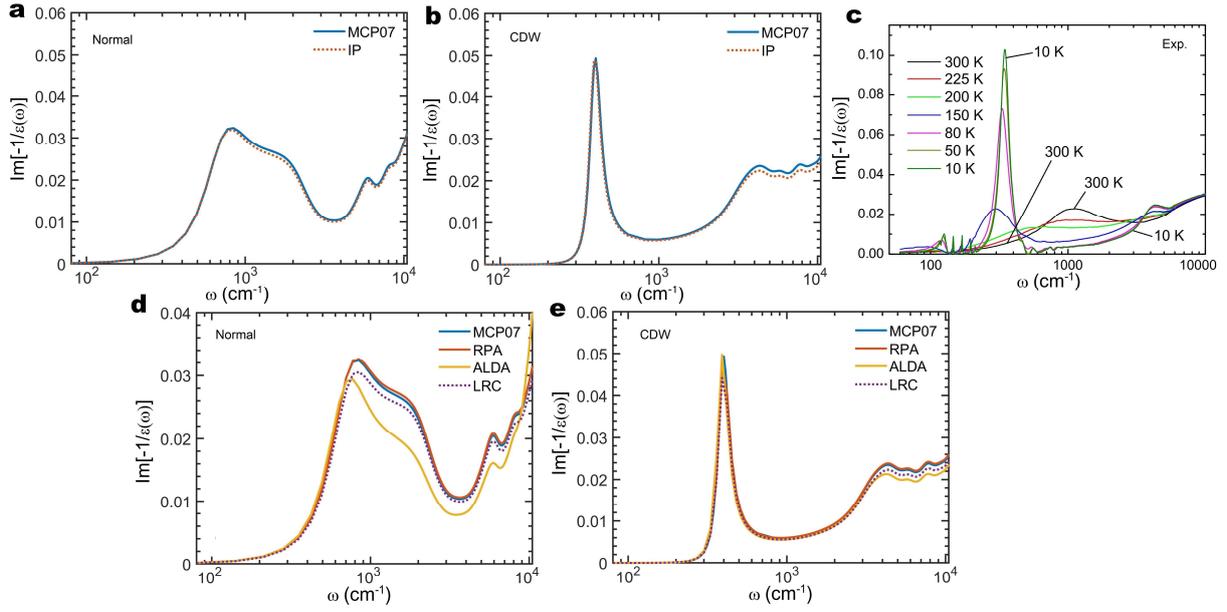

# Supplementary Materials

Table S1. The relaxed lattice constants and *z*-directional distance of Ti and Se atomic layers $d_{Ti-Se}$ in bulk and monolayer $TiSe_2$ with PBE and different meta-GGA approximations. The experimental and hybrid-functional-relaxed lattice constants are listed for comparison. The Grimme D2 correction[1] for the long-range van der Waals interaction has been included in all the functionals, with the global scaling factor of 0.75 appropriate to PBE (but not necessarily to the other functionals in this table).

| Method | Bulk | | | Monolayer | |
|---|---|---|---|---|---|
| | a (Å) | c (Å) | $d_{Ti-Se}$ (Å) | a (Å) | $d_{Ti-Se}$ (Å) |
| Exp.[2–4] | 3.540 | 6.007 | 1.532 | 3.538 | - |
| PBE | 3.503 | 6.162 | 1.555 | 3.498 | 1.563 |
| SCAN | 3.517 | 6.075 | 1.523 | 3.521 | 1.524 |
| r$^2$SCAN | 3.524 | 6.087 | 1.524 | 3.528 | 1.525 |
| MVS | 3.607 | 9.891 | 1.452 | 3.623 | 1.443 |
| TASK | 3.659 | 7.456 | 1.515 | 3.662 | 1.517 |
| HSE06[5] | 3.528 | 6.104 | 1.522 | - | - |
| HSE(17,0)[5] | 3.531 | 6.113 | 1.527 | - | - |

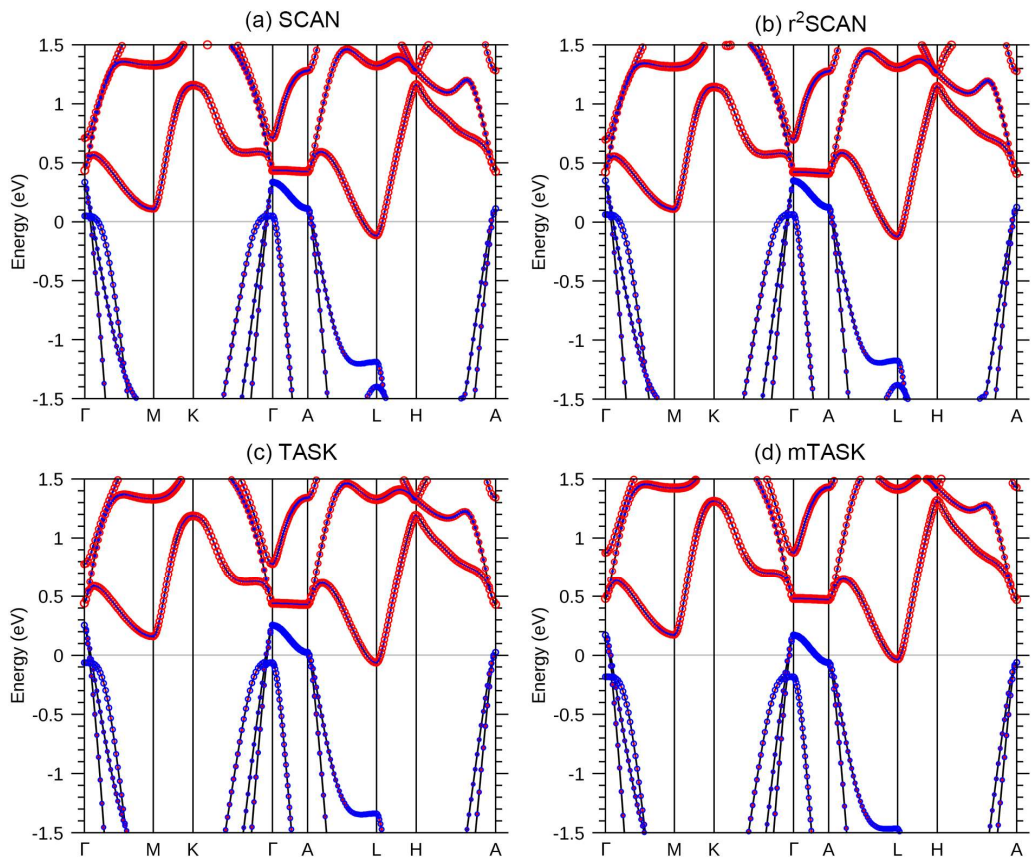

Fig. S1 The atom-characterized band structures of undistorted bulk TiSe$_2$ with different meta-GGA approximations. The red (blue) circles represent the Ti 3$d$ (Se 4$p$) character. The Fermi level is at 0 eV.

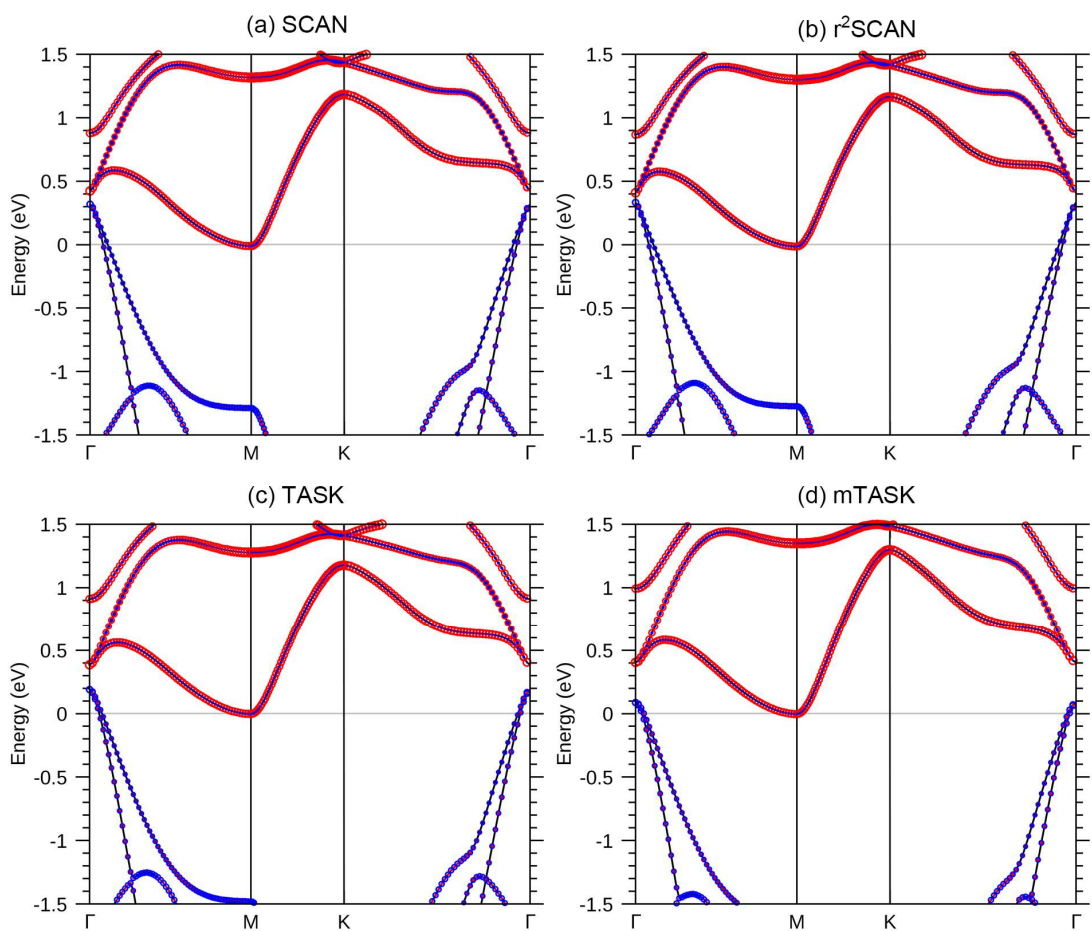

Fig. S2 The atom-characterized band structures of undistorted monolayer TiSe$_2$ with different meta-GGA approximations. The red (blue) circles represent the Ti 3$d$ (Se 4$p$) character. The Fermi level is at 0 eV.

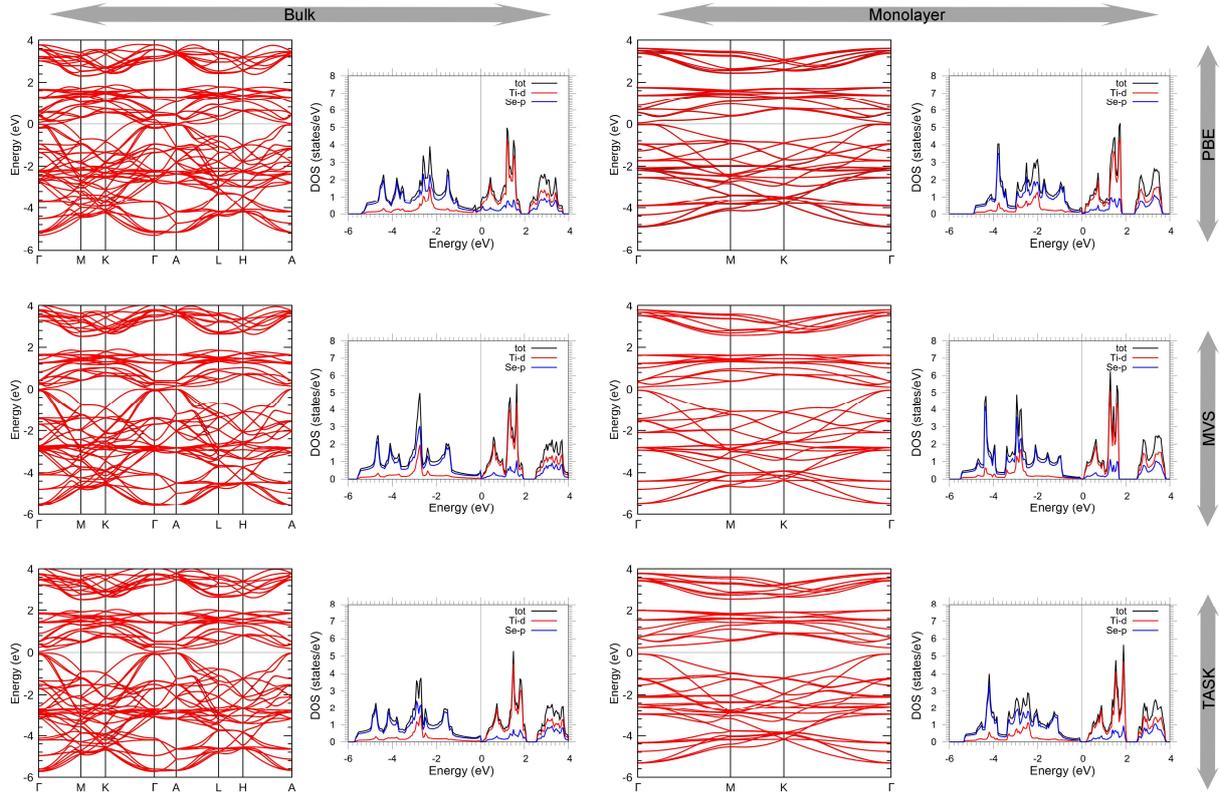

Fig. S3 The band structures and density of states of bulk and monolayer TiSe$_2$ in relaxed CDW supercell with PBE, MVS, and TASK approximations, respectively. The Fermi level is 0 eV. In band structures, the black color denotes the bands calculated by density functional theory, and the red color curves represent the bands from diagonalizing the Wannier function based Hamiltonian.

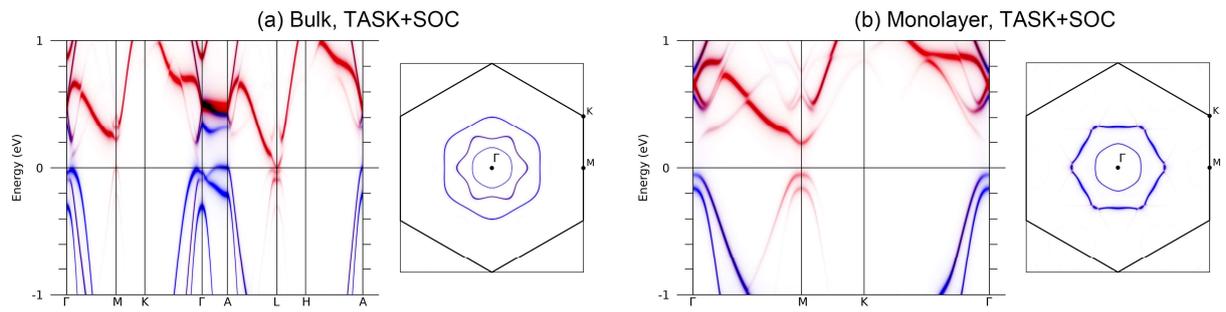

Fig. S4 The unfolded band structures and energy surfaces for (a) bulk and (b) monolayer TiSe$_2$ in relaxed CDW phase with TASK approximation and spin-orbit coupling. The red (blue) color denotes Ti 3$d$ (Se 4$p$) character. The Fermi level $E_F$ is at 0 eV. The displayed energy surface is located at $E_F$-1 eV.

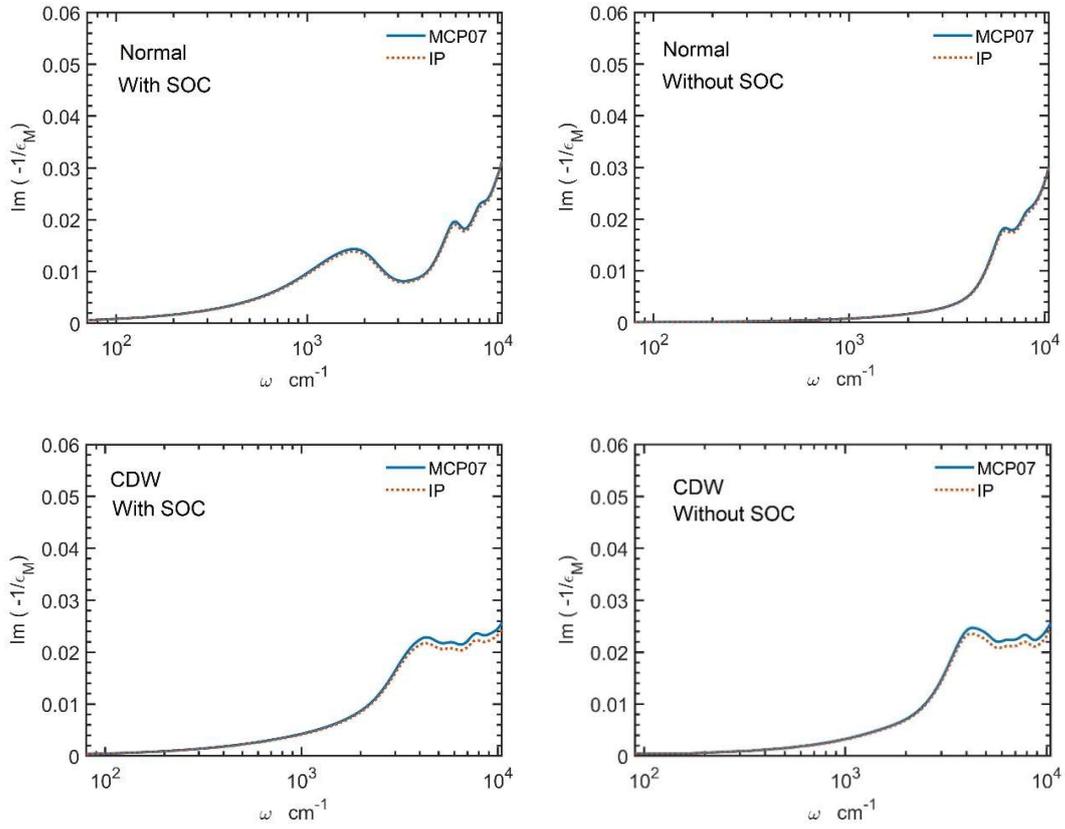

Fig. S5 Comparison of the calculated EELS of 1T TiSe$_2$ with and without the spin-orbit coupling (SOC) effect. Top left: normal phase with SOC; Top right: normal phase without SOC; Bottom left: CDW phase with SOC; Bottom right: CDW phase without SOC.

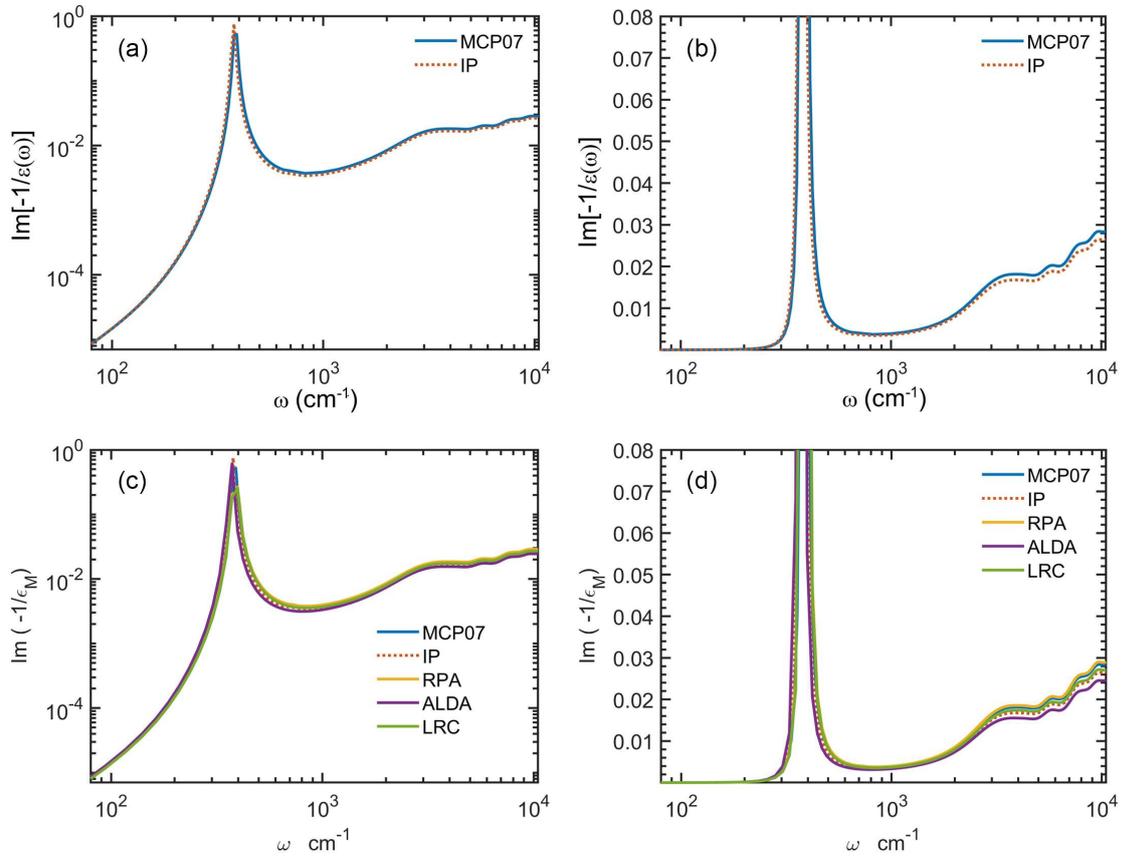

Fig. S6 The EELS for bulk TiSe$_2$ calculated from the gapped CDW ground state with different kernels and the independent particle (IP) approximation. The left plots (a) and (c) are the log-log plots and the right (b) and (d) are the same result with a smaller vertical normal scale.

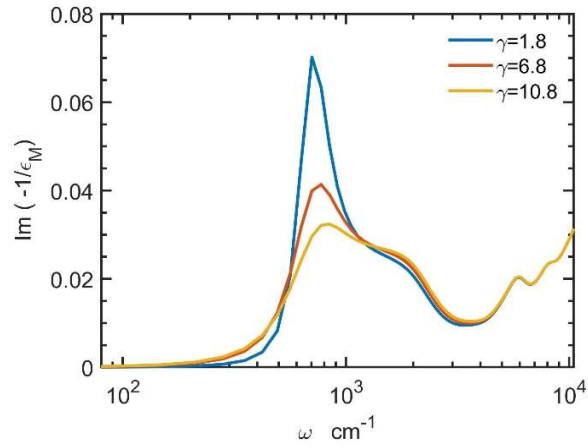

Fig. S7 Calculated EELS curves of 1T TiSe$_2$ at the normal phase with different $\gamma$ values (unit: eV) in the Drude term. The $\omega_{pl}^{unscreened}$=0.5 eV is kept the same.

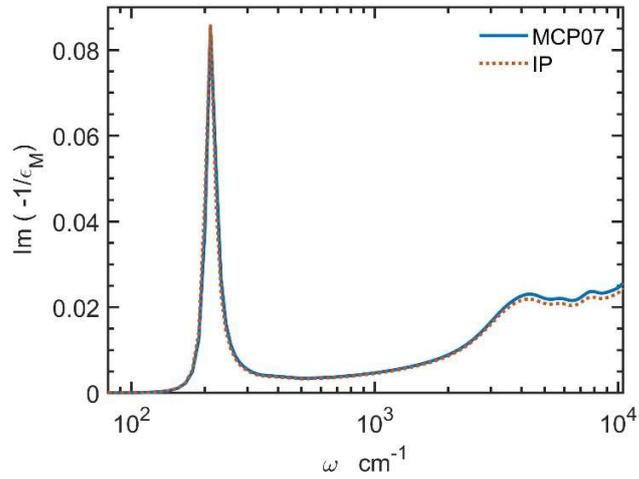

Fig. S8 Calculated EELS of 1T TiSe$_2$ at CDW phase with $\omega_{pl}^{unscreened}$=0.27 and $\gamma$=0.1 eV in the Drude term.